\documentclass{midl} 

\usepackage{amsmath,amssymb,amsfonts}
\usepackage{colortbl}
\usepackage{algorithmic}
\usepackage{graphicx}
\usepackage{algorithm,algorithmic}
\usepackage{multirow}
\usepackage{diagbox}
\usepackage{booktabs}
\usepackage{colortbl}
\usepackage[dvipsnames]{xcolor}

\title[StyCona]{Style Content Decomposition-based Data Augmentation for Domain Generalizable Medical Image Segmentation}

\midlauthor{\Name{Zhiqiang Shen\nametag{$^{1,2,4}$}} \Email{xxszqyy@gmail.com} \\
\Name{Peng Cao\nametag{$^{1,2}$}} \\
\Name{Jinzhu Yang\nametag{$^{1,2}$}} \\
\addr $^{1}$ School of Computer Science and Engineering, Northeastern University, Shenyang, China.\\
\addr $^{2}$ Key Laboratory of Intelligent Computing in Medical Image, Ministry of Education, Northeastern University, Shenyang, China \AND
\Name{Osmar R. Zaiane\nametag{$^{3}$}} \\
\addr $^{3}$ Alberta Machine Intelligence Institute, University of Alberta, Edmonton, Canada \AND
\Name{Zhaolin Chen\nametag{$^{4,5}$}} \\
\addr $^{4}$ Department of Data Science \& AI, Faculty of Information Technology, Monash University, Melbourne, Australia \\
\addr $^{5}$ Monash Biomedical Imaging, Monash University, Melbourne, Australia \\
}

\begin{document}
\maketitle

\begin{abstract}
Due to domain shifts across diverse medical imaging modalities, learned segmentation models often suffer significant performance degradation during deployment. 
We posit that these domain shifts can generally be categorized into two main components:
1) \textbf{"style" shifts}, referring to global disparities in image properties such as illumination, contrast, and color; and 2) \textbf{"content" shifts}, which involve local discrepancies in anatomical structures.
To address the domain shifts in medical image segmentation, we first factorize an image into style codes and content maps, explicitly modeling the "style" and "content" components. 
Building on this, we introduce a \textbf{Sty}le-\textbf{Con}tent decomposition-based data \textbf{a}ugmentation algorithm (StyCona), which performs augmentation on both the global style and local content of source-domain images, enabling the training of a well-generalized model for domain generalizable medical image segmentation.
StyCona is a simple yet effective plug-and-play module that substantially improves model generalization without requiring additional training parameters or modifications to segmentation model architectures.
Experiments on cardiac magnetic resonance imaging and fundus photography segmentation tasks, with single and multiple target domains respectively, demonstrate the effectiveness of StyCona and its superiority over state-of-the-art domain generalization methods.
The code is available at \href{https://github.com/Senyh/StyCona}{\textit{\texttt{https://github.com/Senyh/StyCona}}}.
\end{abstract}


\section{Introduction}
\label{sec:intro}
Medical image segmentation is critical for computer-aided diagnosis. Driven by large and diverse annotated datasets, deep learning-based segmentation models have achieved remarkable progress~\citep{ronneberger2015u,milletari2016v}. However, data distribution mismatches (\emph{a.k.a.}, domain shifts), caused by diverse imaging protocols, equipment vendors, or patient populations, \emph{etc.}, between source (training) and target (testing) domains hinder the generalizability of trained models for clinical deployment~\citep{castro2020causality,zhou2022domain,guan2021domain}. 

\begin{figure*}[!t]
\centering
\includegraphics[width=.8\linewidth]{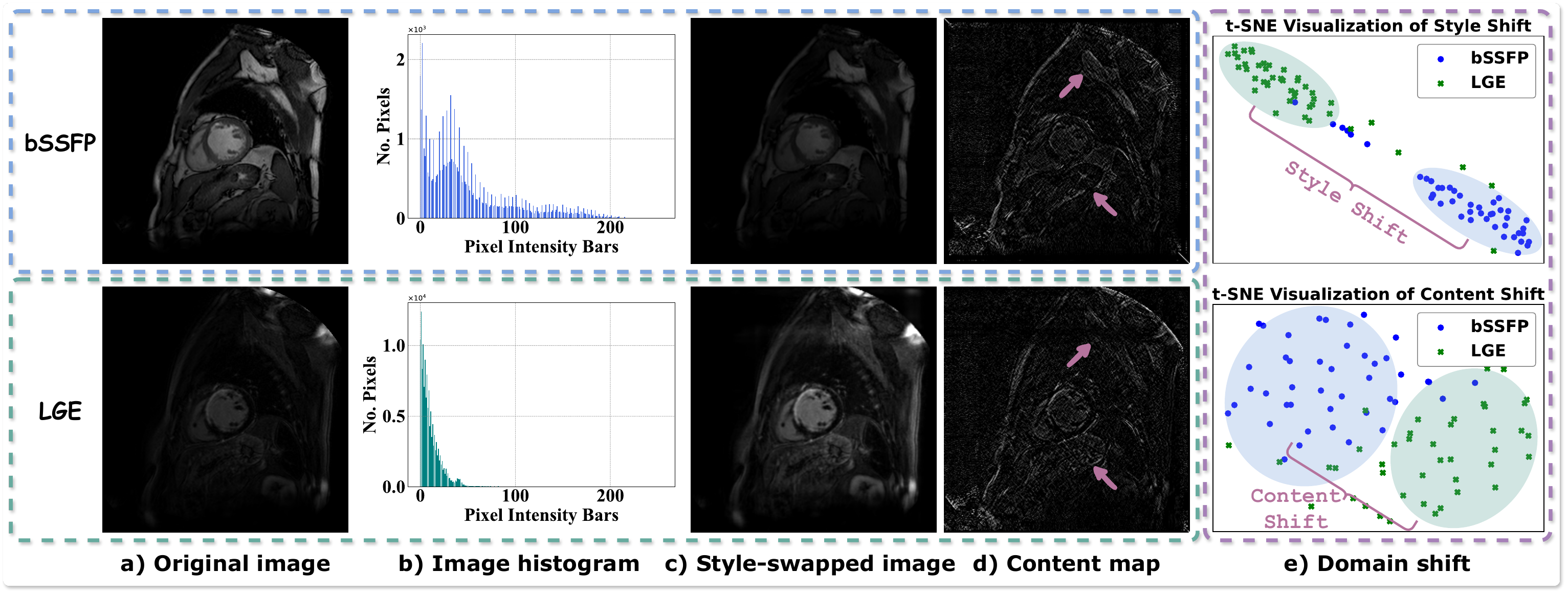}
\caption{Illustration of style codes and content maps for bSSFP and LGE MRI sequences: a) original images showing disparities in both style (global image appearance) and content (local anatomical structures), b) image histograms reflecting domain shifts in terms of pixel intensities, c) style-swapped images generated by swapping the two original images' style codes (top: bSSFP image with LGE style; bottom: LGE image with bSSFP style), d) content maps of the original images, and e) t-SNE~\citep{van2008visualizing} visualization of domain shifts (top: style shift; bottom: content shift).} 
\label{fig:domain_shifts}
\end{figure*}

Domain generalization (DG) aims to mitigate domain shifts by training models on source-domain data and generalizing them to unseen target domains.
Existing DG studies generally promote the learning of domain-invariant representations through two major paradigms: representation learning~\citep{chen2019synergistic,chen2020unsupervised,pei2021disentangle,peng2022unsupervised,gao2022joint,gao2023bayeseg} and data augmentation (\emph{a.k.a.}, domain randomization)~\citep{ouyang2022causality,zhou2021mixstyle,chen2023treasure,chen2025constyx,gu2023cddsa,yang2020fda,xu2021fourier,li2023frequency}. 
The former explicitly explores domain-invariant information via deterministic~\citep{chen2019synergistic,chen2020unsupervised,pei2021disentangle} or statistical~\citep{gao2022joint,gao2023bayeseg} modeling. 
Although such methods yield intuitive decoupled features on source-domain data, their reliance on domain-specific training often hinders the generalization of disentanglement capabilities across different domains.
In contrast, data augmentation-based DG methods alleviate this limitation by expanding the source-domain data distribution and thus implicitly encouraging models to excavate domain-invariant features. 
As a result, this paradigm has become the mainstream approach in the field.
Specifically, these methods typically employ techniques such as Fourier transformation~\citep{yang2020fda,xu2021fourier,li2023frequency}, random convolution~\citep{xu2021robust,ouyang2022causality,choi2023progressive}, and feature statistics editing~\citep{zhou2021mixstyle,chen2023treasure,chen2025constyx}, \emph{etc} to achieve style augmentation on source-domain data.
However, the underlying components involved in medical images remain underexplored.
Modeling these components can provide valuable insights into the nature of domain shifts and guide the design of more effective data augmentation strategies.

To bridge this gap, we introduce a style–content decomposition strategy that decomposes an image into \textit{style codes} and \textit{content maps}, revealing that 1) the style code captures global image characteristics within a given anatomical structure and 2) the content maps describe the image's anatomical structure. 
Fig.~\ref{fig:domain_shifts}(c-d) visualizes style-swapped images (obtained by exchanging the style codes between two images) and content maps. 
It can be observed that the content maps delineate the anatomical structures of the original bSSFP and LGE images, while the style-swapped images exchange appearance between domains but preserve the original anatomical structures.
Furthermore, the shift observed in the style codes is more significant than that in the content maps [Fig.~\ref{fig:domain_shifts}(e)], as global variations in image characteristics are generally greater than local differences in anatomical structures. 
Based on these observations, we categorize domain shifts in medical images into two components: 
1) \textbf{style shifts} (global image property variations) as indicated by deviations in the style codes, 
and 2) \textbf{content shifts} (local anatomical structure discrepancies) as reflected in differences between the content maps. 
Since quantitatively measuring and reducing these two types of shifts is infeasible without access to target domain data during training, we instead perturb the "style" and "content" components of source domain images to generate augmented images from diverse domains (simulating patients undergoing different imaging systems), enabling the training of a well-generalized medical image segmentation model.

To this end, we propose a \textbf{sty}le \textbf{con}tent decomposition-based data \textbf{a}ugmentation algorithm (\textbf{StyCona}) to advance domain generalizable medical image segmentation. 
Specifically, StyCona performs perturbations on the "style" (global image characteristics) and "content" (local anatomical structures) of an image by 1) blending style codes and 2) mixing content maps, respectively.
StyCona generates augmented images with diverse styles and contents while preserving their semantic information, enabling the simulation of images from a wide range of domains for training a well-generalized segmentation model.
We evaluate StyCona on cross-domain cardiac magnetic resonance imaging (MRI) segmentation and optic cup (OC)/optic disk (OD) fundus photography segmentation tasks. Experimental results demonstrate that StyCona is a promising solution for domain generalizable medical image segmentation.

Our main contributions can be summarized as follows:
\begin{itemize}
    \item[$\bullet$] We propose a style–content decomposition strategy to model the underlying components of medical images and provide a deep insight into domain shifts.
    \item[$\bullet$] We propose StyCona, a novel data augmentation algorithm for domain generalizable medical image segmentation. StyCona can be easily integrated into off-the-shelf medical image segmentation backbones and effectively mitigates domain shifts. 
    \item[$\bullet$] StyCona leads to state-of-the-art performance across a wide range of tasks and provides a compelling alternative to Fourier transformation-based, random convolution-based, and feature statistics editing-based domain generalization methods.
\end{itemize}

\begin{figure*}[!t]
\centering
\includegraphics[width=\linewidth]{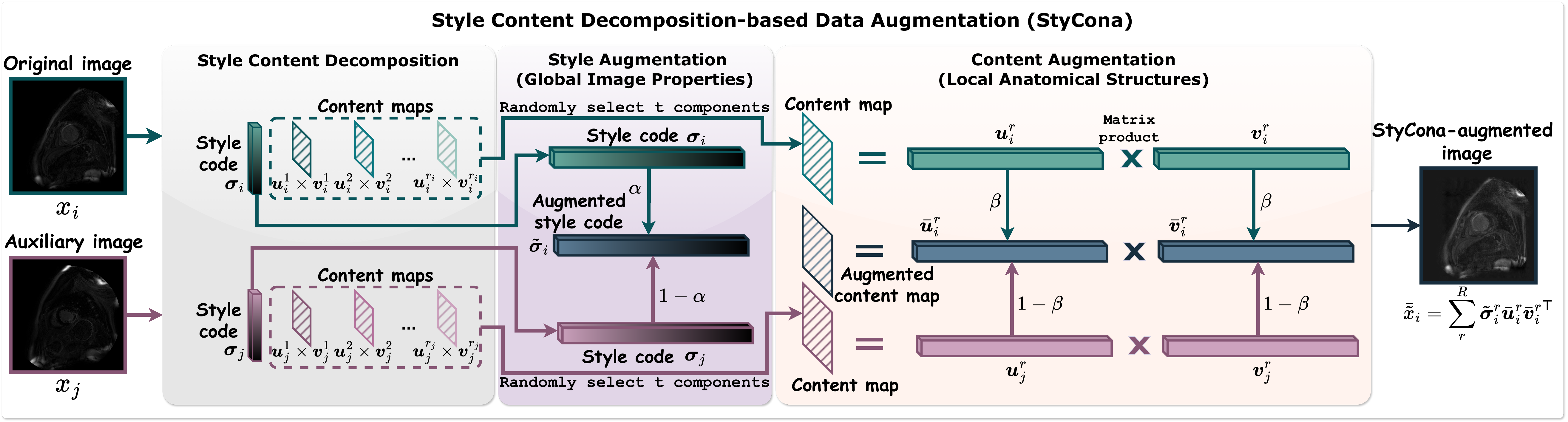}
\caption{Schematic diagram of the proposed \textbf{sty}le \textbf{con}tent decomposition-based data \textbf{a}ugmentation algorithm (\textbf{StyCona}). StyCona includes three steps: 1) style content decomposition, 2) style augmentation, and 3) content augmentation. $x_i$ and $x_j$ represent original images, and $\bar{\tilde{x}}$ denotes a StyCona-augmented sample.} 
\label{fig:stycona}
\end{figure*}

\section{Methodology}
\label{sec:method}
For a domain generalizable medical image segmentation task, the training set can be denoted as $\mathcal{D} = \{\mathcal{S}_k | \mathcal{S}_k = \{(x_{i(k)}, y_{i(k)})_{i=1}^{N_k}\}_{k=1}^K\}$,
where $x_{i(k)}$ represents the $i^{th}$ image in the ${k^{th}}$ domain and $y_{i(k)}$ is its corresponding ground truth segmentation label. 
This paper focuses on the most challenging single-source domain generalization setting (\textit{i.e.}, $K = 1$), in which the segmentation model $f(\cdot; \theta)$ is trained on a single source domain and evaluated on one or more unseen target domains. Note that we omit the domain index for brevity.

\subsection{Style Content Decomposition-based Data Augmentation (StyCona)}
\label{subsec:stycona}
Domain shifts in medical images can be characterized two principal components: 1) \textbf{"style" shifts}, referring to \textit{global} variations in image properties such as illumination, contrast, and color; and 2) \textbf{"content" shifts}, which involve \textit{local} changes in image characteristics and are usually associated with differences in the visibility of specific tissues.
Based on this assumption, we propose StyCona to mitigate both style and content shifts through corresponding style and content augmentations.
As illustrated in Fig.~\ref{fig:stycona}, StyCona mainly includes three steps: 1) style content decomposition, 2) style augmentations, and 3) content augmentation.
Specifically, the style-content decomposition is performed using singular value decomposition (SVD)~\citep{klema1980singular}, which factorizes an image into its \textbf{singular values (style codes)} and \textbf{rank-one matrices (content maps)}.
Then, the style and content augmentations are achieved by perturbing the style codes and content maps, respectively.

\subsubsection{Style Content Decomposition}
Low-rank matrices of an image correspond to its principal components~\cite{abdi2010principal,candes2011robust}. 
Based on this idea, prior work has employed rank-one matrices as the principal parts of an image for restoration tasks~\cite{gao2020rank}. 
We leverage SVD to show that an image’s singular values encode its \textbf{style codes}, while the associated rank-one matrices represent its \textbf{content maps}:
\begin{equation}
    x = \sum^R_{r} \underbrace{\sigma^r}_{style\,code} \, \, \underbrace{\boldsymbol{u}^r {\boldsymbol{v}^r}^\mathsf{T}}_{content\,map}
    \label{eq:svd}
\end{equation}
where 
$R$ denotes the rank of the image, corresponding to the number of non-zero singular values.
SVD provides the optimal rank-one decomposition in the least-squares sense. 
It guarantees that the set $\boldsymbol{\Phi} = \{\boldsymbol{u}^1 {\boldsymbol{v}^1}^\mathsf{T}, \boldsymbol{u}^2 {\boldsymbol{v}^2}^\mathsf{T}, ..., \boldsymbol{u}^R {\boldsymbol{v}^R}^\mathsf{T}\}$ forms a complete basis for the column and row spaces of $x$. 
\begin{itemize}
    \item \textbf{Content Map.} Each rank-one matrix in $\boldsymbol{\Phi}$ functions as a basis pattern of the image, and any image structure can be represented as a linear combination of these patterns. Thus, $\boldsymbol{\Phi}$ completely defines the anatomical structures of the image.
    \item \textbf{Style Code.} Each scalar $\sigma^r$ acts as a global multiplier for the corresponding basis pattern $(\boldsymbol{u}^r {\boldsymbol{v}^r}^\mathsf{T})$.
It modulates the magnitude of pixel intensity in the image, scaling the values of the basis pattern proportionally.
\end{itemize}
As illustrated in Fig.~\ref{fig:domain_shifts}, swapping the style codes of the bSSFP and LGE images results in style transfer between them, while summing all content maps of one of the images reconstructs its complete anatomical structure.

\subsubsection{Style Augmentation} Style augmentation aims to alleviate style shifts (global variations in image properties). As depicted in Fig.~\ref{fig:stycona}, we perform style augmentation on an image $x_i$ by blending its \textbf{style codes} with those of an auxiliary image $x_j$ in an element-wise manner:
\begin{equation}
    \tilde{\sigma}_i^r = \alpha \times \sigma_i^r + (1 - \alpha) \times \sigma_j^r
    \label{eq:style}
\end{equation}
where the weight $\alpha \sim U(0, 1)$ controls the mix strength, $\sigma_i^r$ represents the $r_{th}$ singular value of image $x_i$, and $i/j$ indicates the sample index ($i \neq j$). 

\subsubsection{Content Augmentation} We devise content augmentation to mitigate content shifts (local disparities in anatomical structure).
This is achieved by mixing a set of $t$ randomly selected \textbf{content maps} from an image $x_i$ with another set of $t$ randomly selected content maps from an auxiliary image $x_j$ [Fig.~\ref{fig:stycona}]. The perturbation to each content map is formulated as:
\begin{equation}
    \bar{\boldsymbol{u}}_i^r= \beta \times \boldsymbol{u}_i^r + (1 - \beta) \times \boldsymbol{u}_j^r \quad or \quad  \bar{\boldsymbol{v}}_i^r = \beta \times \boldsymbol{v}_i^r + (1 - \beta) \times \boldsymbol{v}_j^r
    \label{eq:content}
\end{equation}
where the weight $\beta \sim U(0, 1)$ and $\boldsymbol{u}_{i/j}^r$/$\boldsymbol{v}_{i/j}^r$ represents the left/right singular vector of image $x_{i/j}$. 

In a nutshell, StyCona performs style and content augmentation by Eq.~\eqref{eq:svd}\eqref{eq:style}\eqref{eq:content}, and generates an augmented image as $\bar{\tilde{x}}_i = \sum^{R}_{r=1} \tilde{\boldsymbol{\sigma}}_i^r \bar{\boldsymbol{u}}_i^r {\bar{\boldsymbol{v}}_i^{r}}{}^\mathsf{T}$. 

\subsubsection{Loss Supervision} Afterwards, we formulate the segmentation loss $\mathcal{L}$ for training a domain generalizable model $f(\cdot;\theta)$ based on StyCona augmented images:
\begin{equation}
\mathcal{L} = \frac{1}{N}\sum^N_i\mathcal{L}_{seg}(f(\bar{\tilde{x}}_i, \theta), y_i)
\label{eq:seg}
\end{equation}
where $\mathcal{L}_{seg}$ denotes a segmentation criterion.

\section{Experiments and Results}
\label{sec:experiments}

\paragraph{Cross-Domain Cardiac Magnetic Resonance Imaging (MRI) Segmentation} 
The MS-CMR Dataset (\textit{Single Source and Single Target Domain}) dataset~\citep{zhuang2018multivariate} contains 45 subjects, each with bSSFP and LGE MRI sequences, along with ground truth annotations for the right ventricle (RV), left ventricle (LV), and myocardium (MYO). All images are normalized to the range $[0,1]$ and resampled to a uniform resolution of $1.0\times 1.0$ mm. 
This experimental setting evaluates a segmentation model’s ability to generalize across MRI sequences for cardiac structure segmentation, specifically from LGE to bSSFP and vice versa. For each direction, one sequence (\emph{e.g.}, bSSFP) is treated as the source domain and is divided into training and validation sets with a $8:2$ ratio, while another sequence (\emph{e.g.}, LGE) serves as the target domain and is used solely for testing.

\paragraph{Cross-Domain Optic Cup (OC) and Optic Disc (OD) Fundus Image Segmentation}
The Fundus Image Benchmark (\textit{Single Source and Multiple Target Domains})~\citep{chen2023treasure} is established based on the ORIGA~\citep{zhang2010origa}, Drishti-GS~\citep{sivaswamy2014drishti}, REFUGE~\citep{orlando2020refuge}, and RIGA~\citep{almazroa2018retinal} (BinRushed and Magrabia) datasets. 
This setting evaluates the model’s generalization capability for the joint OC and OD segmentation across multiple target domains. The ORIGA dataset is used as the source domain with the training and validation split following TriD~\citep{chen2023treasure}; the other datasets (BinRushed, Drishti-GS, Magrabia, and REFUGE) act as target domains for cross-domain evaluation.

\paragraph{Implementation Details.}
We conducted the experiments using PyTorch~\citep{paszke2019pytorch} on an NVIDIA A40 GPU with 48G GPU memory. 
All compared methods were optimized using an AdamW optimizer~\citep{kingma2014adam} with a fixed learning rate of $1e-4$ during the 100 training epochs for both the cadiac MRI and fundus image segmentation tasks. 
U-Net~\citep{ronneberger2015u} was employed as the baseline segmentation model. 
The combination of cross-entropy and Dice loss~\citep{milletari2016v} was used as the segmentation criterion.
All images were resized to $256 \times 256$ for both training and testing, and the predicted labels were rescaled to their original resolutions for evaluation.
We set the number of perturbed content maps $t = 16$ (please refer to Section~\ref{para:label_invar} for further analysis).

\subsection{Comparison with State of the Arts}
\label{subsec:comparison_with_sota}
We compared StyCona with state-of-the-art DG methods: 
1) Fourier transformation-based: amplitude swap (AmpSwap)~\citep{yang2020fda}, amplitude mixup (AmpMix)~\citep{xu2021fourier}, and FMAug~\citep{li2023frequency}); 
2) Feature statistics editing-based: MixStyle~\citep{zhou2021mixstyle}, TriD~\citep{chen2023treasure}, and ConStyX~\citep{chen2025constyx};
3) Random convolution-based: RandConv~\citep{xu2021robust}, CIDA~\citep{ouyang2022causality}, and PRandConv~\citep{choi2023progressive}. 
We re-implemented all the compared methods in a unified experimental setup for fair comparison.
All the methods are built upon the same baseline segmentation model (U-Net).
In general, StyCona sets the state-of-the-art in all three settings, both quantitatively [Table~\ref{Tab:mscmr} and Table~\ref{Tab:fundus}] and qualitatively [Fig.~\ref{fig:qualitative}], showcasing its superiority in domain generalizable medical image segmentation.

\subsubsection{Results on Cardiac MRI} 
As reported in Table~\ref{Tab:mscmr}, StyCona achieves consistent performance improvements over the compared methods in the cardiac MRI segmentation task. 
In general, almost all the compared methods surpass the baseline model, indicating their effectiveness in addressing domain shifts. 
However, it is worth noting that some methods, \emph{e.g.}, AmpMix~\citep{xu2021fourier} and TriD~\citep{chen2023treasure}, yield segmentation results inferior to those of the baseline model in the bSSFP $\rightarrow$ LGE scenario. 
This may result from the fact that the blended amplitude spectrum in AmpMix and the mixed feature statistics in TriD generate style-augmented images that are insufficient for mitigating content shifts, resulting in the models overfitting spurious correlations between the augmented images and their corresponding segmentation labels.
In contrast, StyCona demonstrates a clear advantage in handling both style and content shifts by style content augmentation. These results validate the key idea of StyCona and highlight its effectiveness in domain generalizable medical image segmentation.

\subsubsection{Results on Fundus Image} 
We further evaluate StyCona in a single-source, multi-target domain setting.
As shown in Table~\ref{Tab:fundus}, all methods generally achieve comparable average performance across the four target domains.
However, the two random convolution-based approaches (\emph{i.e.}, CIDA~\citep{ouyang2022causality} and PRandConv~\citep{choi2023progressive}) yield relatively unsatisfactory results. This may be attributed to that the random convolution operations produce augmented images with distorted content that is inconsistent with the corresponding segmentation labels, thus misleading the segmentation models during training.
In contrast, StyCona shows more robust performance across the four target domains, obtaining the highest average DSC and competitive ASD.
These results further suggest the effectiveness of StyCona in domain generalizable medical image segmentation with multiple target domains.

\begin{table}[!t]
\centering
\caption{Comparison with state-of-the-art methods on the cross-domain cardiac MRI segmentation task (single target domain).
The best and second-best results are highlighted in \textbf{bold} and \underline{underline}, respectively.}
\resizebox{.5\linewidth}{!}{
\begin{tabular}{l|cc|cc}
\toprule[1pt]
\multirow{2}{*}{Method}
& \multicolumn{2}{c|}{bSSFP $\rightarrow$ LGE} & \multicolumn{2}{c}{LGE $\rightarrow$ bSSFP} \\
& DSC $\uparrow$ & ASD $\downarrow$  & DSC $\uparrow$ & ASD $\downarrow$\\ \midrule[1pt]
U-Net~\citep{ronneberger2015u} & 65.97 & 6.55 & 76.91 & 3.41 \\ \midrule 
AmpSwap~\citep{yang2020fda} & 70.93 & 6.48 & 78.74 & 3.26 \\  
AmpMix~\citep{xu2021fourier} & 63.89 & 6.44 & 80.07 & 2.46\\
FMAug~\citep{li2023frequency} & \underline{72.46} & 13.67 & 80.54 & 2.57 \\ \midrule  
MixStyle~\citep{zhou2021mixstyle}  & 69.26 & 10.91 & 80.81 & \underline{2.69} \\ 
TriD~\citep{chen2023treasure}  & 65.12 & 10.25 & 79.14 & 4.60 \\  
ConStyX~\citep{chen2025constyx} & 68.63  & 7.89  &  \underline{81.30}  & 2.85 \\ \midrule 
RandConv~\citep{xu2021robust}& 71.90 & \underline{5.65} & 75.17 & 4.27 \\ 
CIDA~\citep{ouyang2022causality} & 69.72 & 8.63 & 72.85 & 4.61 \\
PRandConv~\citep{choi2023progressive} & 56.61  & 8.75 & 75.57  & 3.38 \\ \midrule 
StyCona (ours) & \textbf{73.39}  & \textbf{4.33}  & \textbf{81.59}  & \textbf{2.24}  \\
\bottomrule[1pt] 
\end{tabular}
}
\label{Tab:mscmr}
\end{table}

\begin{table*}[!t]
\centering
\caption{Comparison with state-of-the-art methods on the fundus image segmentation tasks (Source: ORIGA; Target: REFUGE, Drishti-GS, BinRushed, and Magrabia).
The best and second-best results are highlighted in \textbf{bold} and \underline{underline}, respectively.}
\resizebox{\linewidth}{!}{
\begin{tabular}{l|cc|cc|cc|cc|cc}
\toprule[1pt]
\multirow{2}{*}{Method} & \multicolumn{2}{c|}{BinRushed} & \multicolumn{2}{c|}{Drishti-GS} & \multicolumn{2}{c|}{Magrabia} & \multicolumn{2}{c|}{REFUGE} & \multicolumn{2}{c}{Average} \\ \cline{2-11}
& DSC (\%) $\uparrow$ & ASD $\downarrow$  & DSC (\%) $\uparrow$ & ASD $\downarrow$  & DSC (\%) $\uparrow$ & ASD $\downarrow$  & DSC (\%) $\uparrow$ & ASD $\downarrow$   & DSC (\%) $\uparrow$ & ASD $\downarrow$  \\ \midrule[1pt]
U-Net~\citep{ronneberger2015u}  & 54.82  & 7.07  & 77.99 & 1.16  & 61.19  & 8.10  & 79.49  & 1.89 & 68.37  & 4.56  \\ \midrule 
AmpSwap~\citep{yang2020fda} & 63.56 & 4.36 & 78.16 & 1.13  & \underline{66.13}  & \underline{4.45}  & \underline{81.84}  & \underline{1.45}  & 72.41  & \underline{2.85} \\  
AmpMix~\citep{xu2021fourier} & 65.08 & \underline{4.32} & 77.06 & 1.09  & \textbf{68.14}  & \textbf{4.29}  & 79.50  & 1.67  & 72.44  & \textbf{2.84} \\
FMAug~\citep{li2023frequency} & 63.79 & 5.03  & 78.34 & \underline{1.04}  & 64.33  & 7.28  & \textbf{82.17}  & \textbf{1.29} & 72.16  & 3.66 \\ \midrule  
MixStyle~\citep{zhou2021mixstyle}  & 63.95 & 4.72 & \underline{80.36} & 1.06  & 63.97  & 6.04  & 81.93  & 1.45  & 72.55  & 3.32 \\ 
TriD~\citep{chen2023treasure}  & \underline{66.27} & 4.34 & 77.08 & 1.24  & 65.42  & 4.53  & 79.36  & 1.57  & 72.03  & 2.92 \\  
ConStyX~\citep{chen2025constyx}  & 62.91  & 5.99  & \textbf{80.69}  & \textbf{1.03}  & 64.47  & 7.16  & 80.15  & 1.85  & 72.05  & 4.01  \\ \midrule 
RandConv~\citep{xu2021robust}& 63.75  & 4.70  & 79.35  & 1.13  & 65.61  & 5.07  & 81.52  & 1.46  & \underline{72.56}  & 3.09 \\ 
CIDA~\citep{ouyang2022causality} & 53.16 & 11.34 & 78.55 & 5.18  & 51.27  & 14.61  & 68.09  & 6.62  & 62.76  & 8.44 \\
PRandConv~\citep{choi2023progressive} & 44.60  & 9.84 & 75.23  & 1.48  & 53.34  & 8.68  & 76.61  & 2.57  & 62.44  & 5.64 \\ \midrule 
StyCona (ours) & \textbf{67.80}  & \textbf{3.60}  & 78.43   & 1.08  & 65.71  & 4.37  & 80.68  & 1.49  & \textbf{73.16}  & 2.63  \\
\bottomrule[1pt]
\end{tabular}
}
\label{Tab:fundus}
\end{table*}

\subsubsection{Qualitative results.} 
\begin{figure*}[!t]
\centering
\includegraphics[width=\linewidth]{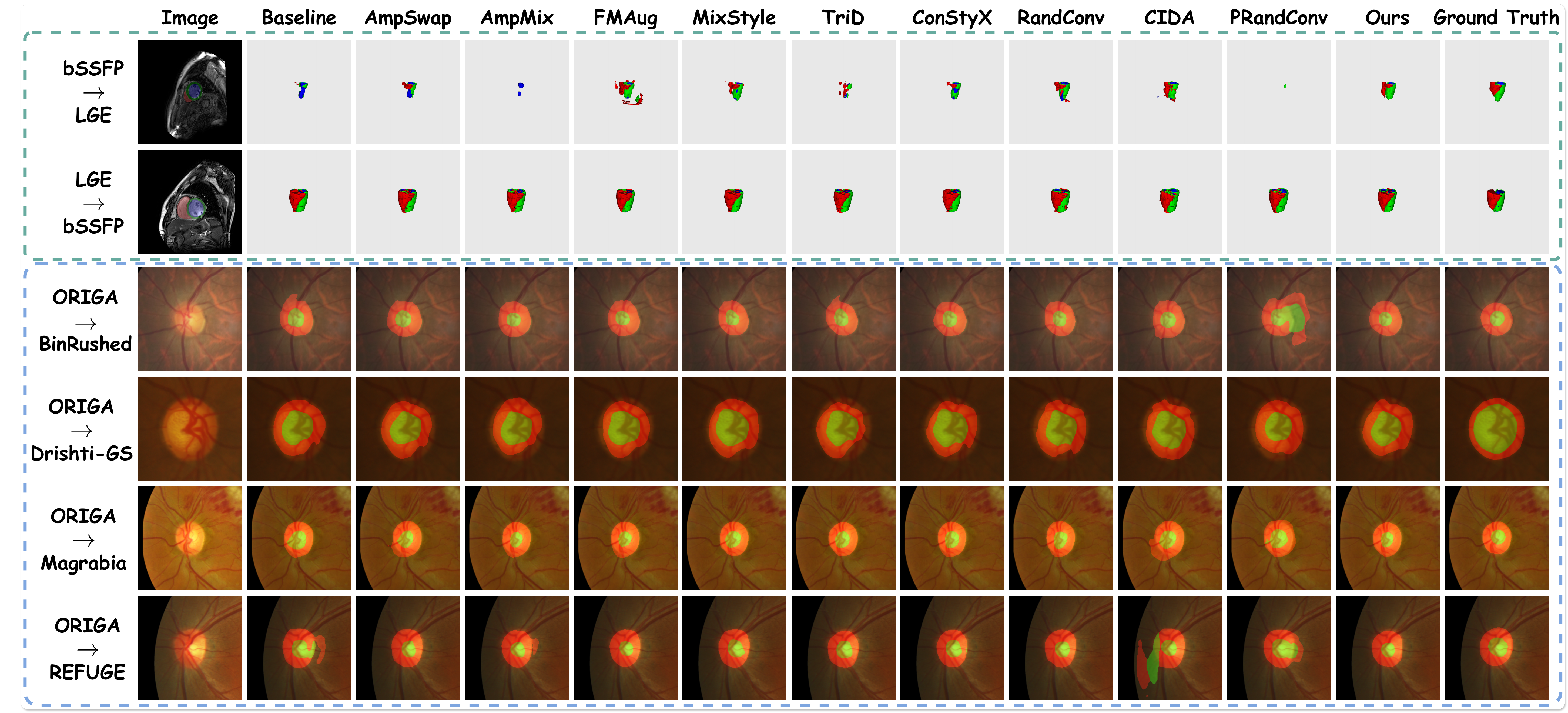}
\caption{Qualitative results on the cardiac MRI segmentation and the OC/OD fundus image segmentation tasks.} 
\label{fig:qualitative}
\end{figure*}

As shown in Fig.~\ref{fig:qualitative}, StyCona produces segmentation results with more precise object delineation. The improvement can be attributed to StyCona's style and content augmentations, which enable the augmented samples to cover a wide range of unseen domains, thereby allowing the trained model to generate accurate segmentation maps for unseen data.
In comparison, methods such as AmpMix~\citep{xu2021fourier}, MixStyle~\citep{zhou2021mixstyle}, and PRandConv~\citep{choi2023progressive} yield relatively unsatisfactory results. AmpMix and MixStyle rely solely on style augmentation by blending amplitude spectra and adjusting the mean and standard deviation of samples, respectively. Meanwhile, the effectiveness of PRandConv is limited by its progressively applied random convolutions, which generate content-distorted images and fail to adequately alleviate variations in anatomical structures.
These qualitative results, consistent with the quantitative performance, further suggest the effectiveness of our approach.

\subsection{Ablation Study}
\label{subsec:ablation_study}
\subsubsection{Effectiveness of each component} In Table \ref{Tab:ablation}, we present ablation experiments to analyze the contributions of each component of StyCona. These experiments were conducted under the cardiac MRI segmentation task. 
The results reveal a consistent trend: segmentation performance improves as the proposed style and content augmentation components are gradually integrated into our method.
Specifically, the baseline model (\emph{i.e.}, U-Net) performs relatively unsatisfactorily on the target domain, as it tends to learn domain-specific decision rules based on source domain data. 
Then, the segmentation performance increases by a large gap when the style augmentation operation is introduced for effectively mitigating style shifts, which are an essential part of domain shifts in these scenarios.
In contrast, the improvement is more significant, with an increase of over 7\% in DSC, when only the content augmentation component is introduced to address the content shifts, another key aspect of domain shifts.
Finally, by incorporating both the style and content augmentation modules to address global variations in image appearance and local differences in anatomical structures, our final method achieves the best performance with DSC of 73.39\% and 81.59\% on the bSSFP $\rightarrow$ LGE and LGE $\rightarrow$ bSSFP scenarios, respectively. 
This result demonstrates the effectiveness of each component and validates our assumptions of the style content decomposition and the corresponding style and content augmentation to alleviate the domain shifts.

\subsubsection{Impact of No. Perturbed Content Maps}
\label{para:label_invar}
\begin{figure*}[!t]
\centering
\includegraphics[width=.9\linewidth]{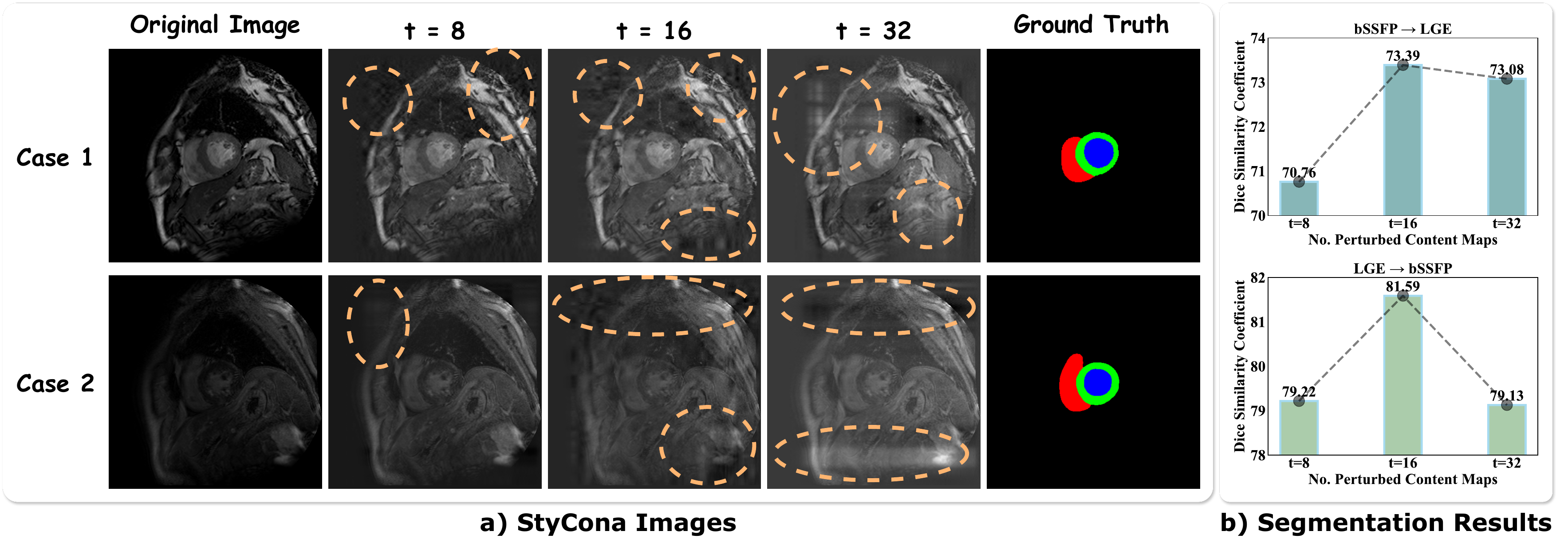} 
\caption{Illustration of a) StyCona-augmented images with different numbers of perturbed content maps ($t=8$, $t=16$, and $t=32$) for two randomly selected original images (The orange dashed circles highlight some content-perturbed regions) and b) Segmentation results for $t=8$, $t=16$, and $t=32$ on cardiac MRI segmentation.
} 
\label{fig:samples}
\end{figure*}

According to Eq.~\eqref{eq:svd}\eqref{eq:style}\eqref{eq:content}, perturbing the style codes alters only an image's style without changing its semantic information corresponding to segmentation labels. 
On the other hand, perturbing $t$ randomly selected content maps results in changes in local anatomical structures (\emph{e.g.}, background tissues or target object-related boundaries), which is the goal of StyCona to enhance the robustness of segmentation models to local anatomical structure variations.
Qualitatively, Fig.~\ref{fig:samples}(a) shows the StyCona-augmented samples with varying numbers of perturbed content maps ($t = 8, 16, 32$). 
Compared with the original images, the augmented images with $t = 8, 16$ exhibit variations in global image appearance and local anatomical structures while remaining consistent with the ground truth segmentation labels. 
In contrast, excessive perturbation is introduced into some target boundaries when $t = 32$, which may alter their ground truth labels. 
Quantitatively, $t = 16$ yields the highest DSC in both the bSSFP $\rightarrow$ LGE and LGE $\rightarrow$ bSSFP scenarios.
Based on these results, we set $t = 16$ in StyCona to balance perturbation strength with the invariance of segmentation labels, ensuring label consistency after content augmentation.

\begin{table*}[!t]
\centering
\caption{Ablation study of StyCona on the cross-sequence setting. 
The best results are highlighted in \textbf{bold}.}
\resizebox{\linewidth}{!}{
\begin{tabular}{c|cc|cc|cc}
\toprule[1pt]
\multirow{2}{*}{Method} & \multicolumn{2}{c|}{StyCona} & \multicolumn{2}{c|}{bSSFP $\rightarrow$ LGE} & \multicolumn{2}{c}{LGE $\rightarrow$ bSSFP} \\
& $\quad$ Style augmentation $\quad$ & $\quad$ Content augmentation $\quad$ & DSC  $\uparrow$ & ASD $\downarrow$ & DSC $\uparrow$ & ASD $\downarrow$ \\
\midrule[1pt]
Baseline	&    &    & 65.97 & 6.55 & 76.91 & 3.41 \\
Baseline + Style &  $\surd$  &  & 68.98  & 3.20  & 80.45  &  2.40 \\
Baseline + Content  &   &  $\surd$  & 72.31  & 5.16  & 81.17  &  2.43\\
Baseline + Style + Content  &  $\surd$  &  $\surd$ & \textbf{73.39} & \textbf{4.33} & \textbf{81.59} & \textbf{2.24} \\  
\bottomrule[1pt]
\end{tabular}
}
\label{Tab:ablation}
\end{table*}

\section{Conclusion}
\label{sec:conclusion}
We propose StyCona, a novel data augmentation algorithm for domain generalizable medical image segmentation. 
Our style content decomposition strategy reveals that an image's singular values function as style codes, governing its global image properties, while its rank-one matrices serve as content maps, determining its anatomical structures.
Based on this decomposition, we categorize domain shifts into style and content shifts \emph{w.r.t.} deviations in style codes and content maps, respectively.
Then, StyCona perturbs style codes for style augmentation and blends content maps for content augmentation, addressing both types of domain shifts accordingly.
Extensive experiments on cardiac MRI segmentation with a single target domain and fundus image segmentation with multiple target domains demonstrate that StyCona is an effective data augmentation technique for domain-generalizable medical image segmentation.

\bibliography{main}

@inproceedings{ronneberger2015u,
  title={U-net: Convolutional networks for biomedical image segmentation},
  author={Ronneberger, Olaf and Fischer, Philipp and Brox, Thomas},
  booktitle={Medical image computing and computer-assisted intervention--MICCAI 2015: 18th international conference, Munich, Germany, October 5-9, 2015, proceedings, part III 18},
  pages={234--241},
  year={2015},
  organization={Springer}
}

@inproceedings{milletari2016v,
  title={V-net: Fully convolutional neural networks for volumetric medical image segmentation},
  author={Milletari, Fausto and Navab, Nassir and Ahmadi, Seyed-Ahmad},
  booktitle={2016 fourth international conference on 3D vision (3DV)},
  pages={565--571},
  year={2016},
  organization={Ieee}
}

@article{gao2023bayeseg,
  title={BayeSeg: Bayesian modeling for medical image segmentation with interpretable generalizability},
  author={Gao, Shangqi and Zhou, Hangqi and Gao, Yibo and Zhuang, Xiahai},
  journal={Medical Image Analysis},
  volume={89},
  pages={102889},
  year={2023},
  publisher={Elsevier}
}

@inproceedings{gao2022joint,
  title={Joint modeling of image and label statistics for enhancing model generalizability of medical image segmentation},
  author={Gao, Shangqi and Zhou, Hangqi and Gao, Yibo and Zhuang, Xiahai},
  booktitle={International Conference on Medical Image Computing and Computer-Assisted Intervention},
  pages={360--369},
  year={2022},
  organization={Springer}
}

@inproceedings{chen2019synergistic,
  title={Synergistic image and feature adaptation: Towards cross-modality domain adaptation for medical image segmentation},
  author={Chen, Cheng and Dou, Qi and Chen, Hao and Qin, Jing and Heng, Pheng-Ann},
  booktitle={Proceedings of the AAAI conference on artificial intelligence},
  volume={33},
  number={01},
  pages={865--872},
  year={2019}
}

@article{chen2020unsupervised,
  title={Unsupervised bidirectional cross-modality adaptation via deeply synergistic image and feature alignment for medical image segmentation},
  author={Chen, Cheng and Dou, Qi and Chen, Hao and Qin, Jing and Heng, Pheng Ann},
  journal={IEEE transactions on medical imaging},
  volume={39},
  number={7},
  pages={2494--2505},
  year={2020},
  publisher={IEEE}
}

@inproceedings{peng2022unsupervised,
  title={Unsupervised domain adaptation for cross-modality retinal vessel segmentation via disentangling representation style transfer and collaborative consistency learning},
  author={Peng, Linkai and Lin, Li and Cheng, Pujin and Huang, Ziqi and Tang, Xiaoying},
  booktitle={2022 IEEE 19th International Symposium on Biomedical Imaging (ISBI)},
  pages={1--5},
  year={2022},
  organization={IEEE}
}

@article{gu2023cddsa,
  title={CDDSA: Contrastive domain disentanglement and style augmentation for generalizable medical image segmentation},
  author={Gu, Ran and Wang, Guotai and Lu, Jiangshan and Zhang, Jingyang and Lei, Wenhui and Chen, Yinan and Liao, Wenjun and Zhang, Shichuan and Li, Kang and Metaxas, Dimitris N and others},
  journal={Medical Image Analysis},
  volume={89},
  pages={102904},
  year={2023},
  publisher={Elsevier}
}

@article{castro2020causality,
  title={Causality matters in medical imaging},
  author={Castro, Daniel C and Walker, Ian and Glocker, Ben},
  journal={Nature Communications},
  volume={11},
  number={1},
  pages={3673},
  year={2020},
  publisher={Nature Publishing Group UK London}
}

@inproceedings{yang2020fda,
  title={Fda: Fourier domain adaptation for semantic segmentation},
  author={Yang, Yanchao and Soatto, Stefano},
  booktitle={Proceedings of the IEEE/CVF conference on computer vision and pattern recognition},
  pages={4085--4095},
  year={2020}
}

@inproceedings{xu2021fourier,
  title={A fourier-based framework for domain generalization},
  author={Xu, Qinwei and Zhang, Ruipeng and Zhang, Ya and Wang, Yanfeng and Tian, Qi},
  booktitle={Proceedings of the IEEE/CVF Conference on Computer Vision and Pattern Recognition},
  pages={14383--14392},
  year={2021}
}

@inproceedings{li2023frequency,
  title={Frequency-mixed single-source domain generalization for medical image segmentation},
  author={Li, Heng and Li, Haojin and Zhao, Wei and Fu, Huazhu and Su, Xiuyun and Hu, Yan and Liu, Jiang},
  booktitle={International Conference on Medical Image Computing and Computer-Assisted Intervention},
  pages={127--136},
  year={2023},
  organization={Springer}
}

@inproceedings{zhou2021mixstyle,
  title={Domain Generalization with MixStyle},
  author={Zhou, Kaiyang and Yang, Yongxin and Qiao, Yu and Xiang, Tao},
  booktitle={International Conference on Learning Representations},
  year={2021}
}

@inproceedings{chen2023treasure,
  title={Treasure in distribution: a domain randomization based multi-source domain generalization for 2d medical image segmentation},
  author={Chen, Ziyang and Pan, Yongsheng and Ye, Yiwen and Cui, Hengfei and Xia, Yong},
  booktitle={International Conference on Medical Image Computing and Computer-Assisted Intervention},
  pages={89--99},
  year={2023},
  organization={Springer}
}

@inproceedings{chen2025constyx,
  title={ConStyX: Content Style Augmentation for Generalizable Medical Image Segmentation},
  author={Chen, Xi and Shen, Zhiqiang and Cao, Peng and Yang, Jinzhu and Zaiane, Osmar R},
  booktitle={International Conference on Medical Image Computing and Computer-Assisted Intervention},
  pages={100--110},
  year={2025},
  organization={Springer}
}

@inproceedings{xu2021robust,
  title={Robust and Generalizable Visual Representation Learning via Random Convolutions},
  author={Zhenlin Xu and Deyi Liu and Junlin Yang and Colin Raffel and Marc Niethammer},
  booktitle={International Conference on Learning Representations},
  year={2021}
}

@inproceedings{choi2023progressive,
  title={Progressive random convolutions for single domain generalization},
  author={Choi, Seokeon and Das, Debasmit and Choi, Sungha and Yang, Seunghan and Park, Hyunsin and Yun, Sungrack},
  booktitle={Proceedings of the IEEE/CVF Conference on Computer Vision and Pattern Recognition},
  pages={10312--10322},
  year={2023}
}

@article{ouyang2022causality,
  title={Causality-inspired single-source domain generalization for medical image segmentation},
  author={Ouyang, Cheng and Chen, Chen and Li, Surui and Li, Zeju and Qin, Chen and Bai, Wenjia and Rueckert, Daniel},
  journal={IEEE Transactions on Medical Imaging},
  volume={42},
  number={4},
  pages={1095--1106},
  year={2022},
  publisher={IEEE}
}

@article{zhuang2018multivariate,
  title={Multivariate mixture model for myocardial segmentation combining multi-source images},
  author={Zhuang, Xiahai},
  journal={IEEE transactions on pattern analysis and machine intelligence},
  volume={41},
  number={12},
  pages={2933--2946},
  year={2018},
  publisher={IEEE}
}

@article{paszke2019pytorch,
  title={Pytorch: An imperative style, high-performance deep learning library},
  author={Paszke, Adam and Gross, Sam and Massa, Francisco and Lerer, Adam and Bradbury, James and Chanan, Gregory and Killeen, Trevor and Lin, Zeming and Gimelshein, Natalia and Antiga, Luca and others},
  journal={Advances in neural information processing systems},
  volume={32},
  year={2019}
}

@inproceedings{kingma2014adam,
  title={Adam: A method for stochastic optimization},
  author={Kingma, Diederik P and Ba, Jimmy},
  booktitle={International Conference on Learning Representations},
  year={2015}
}

@article{klema1980singular,
  title={The singular value decomposition: Its computation and some applications},
  author={Klema, Virginia and Laub, Alan},
  journal={IEEE Transactions on automatic control},
  volume={25},
  number={2},
  pages={164--176},
  year={1980},
  publisher={IEEE}
}

@article{zhou2022domain,
  title={Domain generalization: A survey},
  author={Zhou, Kaiyang and Liu, Ziwei and Qiao, Yu and Xiang, Tao and Loy, Chen Change},
  journal={IEEE Transactions on Pattern Analysis and Machine Intelligence},
  volume={45},
  number={4},
  pages={4396--4415},
  year={2022},
  publisher={IEEE}
}

@article{guan2021domain,
  title={Domain adaptation for medical image analysis: a survey},
  author={Guan, Hao and Liu, Mingxia},
  journal={IEEE Transactions on Biomedical Engineering},
  volume={69},
  number={3},
  pages={1173--1185},
  year={2021},
  publisher={IEEE}
}

@article{van2008visualizing,
  title={Visualizing data using t-SNE.},
  author={Van der Maaten, Laurens and Hinton, Geoffrey},
  journal={Journal of machine learning research},
  volume={9},
  number={11},
  year={2008}
}

@article{abdi2010principal,
  title={Principal component analysis},
  author={Abdi, Herv{\'e} and Williams, Lynne J},
  journal={Wiley interdisciplinary reviews: computational statistics},
  volume={2},
  number={4},
  pages={433--459},
  year={2010},
  publisher={Wiley Online Library}
}

@article{candes2011robust,
  title={Robust principal component analysis?},
  author={Cand{\`e}s, Emmanuel J and Li, Xiaodong and Ma, Yi and Wright, John},
  journal={Journal of the ACM (JACM)},
  volume={58},
  number={3},
  pages={1--37},
  year={2011},
  publisher={ACM New York, NY, USA}
}

@article{pei2021disentangle,
  title={Disentangle domain features for cross-modality cardiac image segmentation},
  author={Pei, Chenhao and Wu, Fuping and Huang, Liqin and Zhuang, Xiahai},
  journal={Medical Image Analysis},
  volume={71},
  pages={102078},
  year={2021},
  publisher={Elsevier}
}

@article{gao2020rank,
  title={Rank-one network: An effective framework for image restoration},
  author={Gao, Shangqi and Zhuang, Xiahai},
  journal={IEEE Transactions on Pattern Analysis and Machine Intelligence},
  volume={44},
  number={6},
  pages={3224--3238},
  year={2020},
  publisher={IEEE}
}

@article{orlando2020refuge,
  title={Refuge challenge: A unified framework for evaluating automated methods for glaucoma assessment from fundus photographs},
  author={Orlando, Jos{\'e} Ignacio and Fu, Huazhu and Breda, Jo{\~a}o Barbosa and Van Keer, Karel and Bathula, Deepti R and Diaz-Pinto, Andr{\'e}s and Fang, Ruogu and Heng, Pheng-Ann and Kim, Jeyoung and Lee, JoonHo and others},
  journal={Medical image analysis},
  volume={59},
  pages={101570},
  year={2020},
  publisher={Elsevier}
}

@inproceedings{zhang2010origa,
  title={Origa-light: An online retinal fundus image database for glaucoma analysis and research},
  author={Zhang, Zhuo and Yin, Feng Shou and Liu, Jiang and Wong, Wing Kee and Tan, Ngan Meng and Lee, Beng Hai and Cheng, Jun and Wong, Tien Yin},
  booktitle={2010 Annual international conference of the IEEE engineering in medicine and biology},
  pages={3065--3068},
  year={2010},
  organization={IEEE}
}

@inproceedings{sivaswamy2014drishti,
  title={Drishti-gs: Retinal image dataset for optic nerve head (onh) segmentation},
  author={Sivaswamy, Jayanthi and Krishnadas, SR and Joshi, Gopal Datt and Jain, Madhulika and Tabish, A Ujjwaft Syed},
  booktitle={2014 IEEE 11th international symposium on biomedical imaging (ISBI)},
  pages={53--56},
  year={2014},
  organization={IEEE}
}

@inproceedings{almazroa2018retinal,
  title={Retinal fundus images for glaucoma analysis: the RIGA dataset},
  author={Almazroa, Ahmed and Alodhayb, Sami and Osman, Essameldin and Ramadan, Eslam and Hummadi, Mohammed and Dlaim, Mohammed and Alkatee, Muhannad and Raahemifar, Kaamran and Lakshminarayanan, Vasudevan},
  booktitle={Medical Imaging 2018: Imaging Informatics for Healthcare, Research, and Applications},
  volume={10579},
  pages={55--62},
  year={2018},
  organization={SPIE}
}

\end{document}